\documentstyle[preprint,aps]{revtex}
\input psfig

\begin{document}

\bigskip

\bigskip

\ \ \ \ \ \ \ \ \ \ \ \ \ \ \ ''COMPOSITE PARTICLES'' and the EIGENSTATES of

\bigskip

\ \ \ \ \ \ \ \ \ \ CALOGERO-SUTHERLAND and RUIJSENAARS-SCHNEIDER

\ \ \ \ \ \ \ \ \ \ \ \ \ \ \ \ \ \ \ \ \ \ \ \ \ \ \ \ \ \ \ \ \ \ \ \ \ 

\ \ \ \ \ \ \ \ \ \ \ \ \ \ \ \ \ \ \ \ \ \ \ \ \ \ \ \ \ \ \ \ \ \ \ \ \ \
\ \ \ \ \ \ \ \ \ \ \ \ by\ \ \ \ \ \ \ \ \ \ \ \ \ \ \ \ \bigskip

\ \ \ \ \ \ \ \ \ \ \ \ \ \ \ \ \ \ \ \ \ \ \ \ \ \ \ \ \ \ \ \ \ \ \ \ \ \
\ \ \ \ \ M. C. Berg\`{e}re

\bigskip

\ \ \ \ \ \ \ \ \ \ \ \ \ \ \ \ \ \ \ \ \ \ \ Service de Physique
Th\'{e}orique, CEA-Saclay

\ \ \ \ \ \ \ \ \ \ \ \ \ \ \ \ \ \ \ \ \ \ \ \ \ \ \ \ F-91191 Gif sur
Yvette cedex, France

\bigskip

\ \ \ \ \ \ \ \ \ \ \ \ \ \ \ \ \ \ \ \ \ \ \ \ \ \ \ \ \ \ email:
bergere@spht.saclay.cea.fr

\bigskip

\bigskip

{\bf Abstract: }We establish a one-to-one correspondance between the
''composite particles'' with $N$ particles and the Young tableaux with at
most $N$ rows. We apply this correspondance to the models of
Calogero-Sutherland and Ruijsenaars-Schneider and we obtain a momentum space
representation of the ''composite particles'' in terms of creation operators
attached to the Young tableaux. Using the technique of bosonisation, we
obtain a position space representation of the ''composite particles'' in
terms of products of vertex operators. In the special case where the
''composite particles'' are bosons and if we add one extra quasiparticle or
quasihole, we construct the ground state wave functions corresponding to the
Jain series $\nu =p/\left( 2np\pm 1\right) $ of the fractional quantum Hall
effect.

\bigskip

\bigskip

\bigskip

\section{Introduction}

\bigskip

In a recent publication$^{\left[ 1\right] }$, we introduced the concept of
''composite particles'' as a quasi-geometric construction which distributes $%
N=pm+r\;\;\;(0<r\leq m)$\ \ particles over a set of $d$ states. The
particles are decomposed into $p$ groups of $m$ particles each called
''composite particles'', and the remaining $r$ particles are called
uncomplete ''composite particles''.

The rule which distributes the $N$ particles over the $d$ states are such
that the number of possible configurations is 
\begin{equation}
C_{d+N-1-lE\left( \frac{N-1}{m}\right) }^{N}  \label{1}
\end{equation}
where $E\left( x\right) $ means integer part of $x$ and where $l\geq 0$ and $%
m>0$ are two integers. We note that if $l=0$ we obtain the statistic for the
bosons, and if $l=m=1$ we have the statistic for the fermions. When the
number $N$ of particles is large, the function integer part in $E\left( 
\frac{N-1}{m}\right) $ becomes diluted and the number of configurations is
given by Haldane's formula$^{\left[ 2\right] }$%
\begin{equation}
C_{d+\left( 1-g\right) \left( N-1\right) }^{N}  \label{2}
\end{equation}
where the fraction $g=\frac{l}{m}.\;$In that sense, the ''composite
particles'' reproduce the fractional statistic introduced by Haldane and
extend it naturally for any finite $N\;$and finite $d;$ in addition it gives
a geometrical interpretation of this statistic.

A ''composite particle''$\;$(fig.1) is a set of $m$ particles and $d\geq l$
states with the constraint that the $(l-1)$ bottom states are empty and the $%
l^{th}$ state (from the bottom) has at least one particle. We define an
uncomplete ''composite particle'' as a set of $\;0<r\leq m$ \ particles with
no constraint on the empty states.

The main constraint in this construction is that a non empty state is
entirely included inside a ''composite particle''. The configurations where
a non empty state should be splitted into two consecutive ''composite
particles'' are forbidden (fig.1).

\bigskip

The partition function for the ''composite particles'' when all the states
have the same energy $E$ has been calculated somewhere else$^{\left[ 3\right]
};$we reproduced the well-known result$^{\left[ 4-8\right] }$ for the
average number of particles per state in the thermodynamic limit

\bigskip 
\begin{equation}
n\left( y,g\right) =\frac{1}{W\left( y,g\right) +g}  \label{3}
\end{equation}
where $y=\exp \left( -E/kT\right) $ and where the function $W\left(
y,g\right) $ satisfies the equation 
\begin{equation}
\left[ W\left( y,g\right) \right] ^{g}\left[ 1+W\left( y,g\right) \right]
^{1-g}=y^{-1}  \label{4}
\end{equation}
When the states\ have a linear distribution of energy so that the state $i$
contributes to the partition function by a factor 
\begin{equation}
x^{i-1}y=\exp \left[ -\left( E+\left( i-1\right) v\right) /kT\right]
\label{5}
\end{equation}
we proved in ref.$\left[ 3\right] $ that the partition function, in the
thermodynamic limit ($d\rightarrow \infty ,x\rightarrow 1),$ is 
\begin{equation}
Z\sim \exp \left( \frac{d}{\ln \xi ^{-1}}\int_{\xi y}^{y}\frac{du}{u}\ln %
\left[ 1+W^{-1}\left( u,g\right) \right] \right)  \label{6}
\end{equation}
where $x=\xi ^{\frac{1}{d}}$ and where the above integral is transformed in
the litterature$^{\left[ 9-12\right] }$ into a Rogers dilogarithm function.
In ref.$\left[ 3\right] $, we transformed the partition function into a
large $d$ expansion and consequently, we determined the finite $d-$size
corrections to (6). These results show essentially that our construction of
the ''composite particles'' reproduces perfectly well previous results by
Hikami$^{\left[ 9\right] }$ (effective central charge for a $g$-on gas with
fractional exclusion statistic), by van Elburg and Schoutens$^{\left[ 10%
\right] }$ (quasi-particles in fractional quantum Hall effect edge theories),%
$\;$by Kedem, Klassen, McCoy and Melzer$^{\left[ 11\right] }\;$in the
context of conformal field theory.\ This partition function is called
the''universal chiral partition function for exclusion statistic'' by
Berkovitch and McCoy$^{\left[ 12\right] }.$

\bigskip

Since we believe that our definition of ''composite particles'' is the basic
construction which generates fractional statistic, it seems necessary to
establish the link between ''composite particles'' and the eigenstates of
the Hamiltonian for the models of Calogero-Sutherland$^{\left[ 5,13\right] }$
and Ruijsenaars-Schneider$^{\left[ 14\right] }$ which are known to satisfy
fractional statistic. It is the purpose of this publication to establish
this correspondance and to study some of its consequences. From the complete
knowledge of the eigenfunctions of these two models, we have beeen able to
construct a momentum space representation of the ''composite particles'' in
terms of creation operators attached to the corresponding Young tableaux;
also, using the well-known bosonisation procedure in terms of products of
vertex operators, we have defined a position space representation of the
''composite particles''.

As an application, we have constructed the ground state wave functions for
the electron, the quasiparticle and the quasihole in a fractional quantum
Hall effect. To achieve this goal, we apply the vertex operator formalism to
the case where the $p$ ''composite particles'' are bosons and made of $m$
(even) quasiparticles while the uncomplete ''composite particle'' is either
a quasiparticle or a quasihole $(r=1).$ When we have only one ''composite
particle''$\;(p=1)$ we obtain for the ground state wave function the
Laughlin wave function$^{\left[ 15\right] };$ more generally, for $p$
''composite particles'', we obtain the ground state wave function as the
product of a boson wave function with satisfies the discrete $Z_{p}$
symmetry times the usual fermionic wave function for the extra quasiparticle
or quasihole. The corresponding filling factor defined as the number of
''composite particles'' divided by the total number of quasiparticles minus
the total number of quasiholes ($\nu =\frac{p}{N_{\pm }})$ becomes, in this
special case where $m=2n$ \ is even,$\allowbreak $ 
\begin{equation}
\nu =\frac{p}{2np\pm 1}  \label{7}
\end{equation}
This filling factor defines the Jain series$^{\left[ 16\right] }$ which
characterise in the fractional quantum Hall effect \bigskip most of the
different plateaus in the Hall resistance.

In section 2, we establish a one-to-one correspondance between the
configurations of the ''composite particles'' with $N$ particles and the
Young tableaux with at most $N$ rows. Each particle corresponds to one row
of the Young tableau; the length of the row is the difference between the
''momentum'' which labels the state where the particle is located in the
given configuration and the ''momentum'' which labels the state where this
same particle is located in the ground state.

Then, in order to construct the link with the models of Calogero-Sutherland
(Ruijsenaars-Schneider), we introduce the notion of shifted momentum (which
is a consequence of the thermodynamic Bethe Ansatz$^{\left[ 17\right] }$)
where the ''momentum'' of the particles inside a given ''composite
particle'' are successively shifted by $0,\frac{l}{m,}\frac{2l}{m},...,\frac{%
\left( m-1\right) l}{m}.$ As a result, we obtain the momentum (rapidities)
for the eigenstates of the Hamiltonian which describes the model of
Calogero-Sutherland (Ruijsenaars-Schneider) up to a global shift which comes
from the labelling convention of the momenta in the Fermi sea.

\bigskip

This correspondance being established, we remind in section 3, the formalism
of bosonisation which has been developped by many authors$^{\left[ 18\right]
}$ and which is based on the algebra of the vertex operators 
\begin{equation}
V\left( z\right) =e^{\sqrt{\beta }Q}\exp \left( \sum_{n>0}\sqrt{\frac{1-t^{n}%
}{1-q^{n}}}\frac{a_{n}^{+}}{n}z^{n}\right) \;z^{\sqrt{\beta }a_{0}}\exp
\left( -\sum_{n>0}\sqrt{\frac{1-t^{n}}{1-q^{n}}}\frac{a_{n}}{n}z^{-n}\right)
\label{8}
\end{equation}
where $t=q^{\beta },$ where the operators $a_{n}$ and $a_{n}^{+}$ satisfy
the commutation relations 
\begin{equation}
\left[ a_{n},a_{n^{\prime }}^{+}\right] =n\;\delta _{n,n^{\prime }}
\label{9}
\end{equation}
and where the operators $a_{0}$ and $Q$ satisfy 
\begin{equation}
\left[ a_{0},Q\right] =1  \label{10}
\end{equation}
all other commutators being nul. In (8), $t$ and $q$ are two parameters
which characterise the homogeneous, symmetric Macdonald polynomials$^{\left[
19\right] }$ of several variables. The product of these vertex operators at
different values of $z$ generates the wave functions for the eigenstates of
the model of Ruijsenaars-Schneider. In the limit $t=q^{\beta }$ and $%
q\rightarrow 1,$ the Macdonald polynomials become the Jack polynomials and
we generate the wave functions for the eigenstates of the model of
Calogero-Sutherland.

The product of several vertex operators $V_{+}\left( z\right) $ \ (where $+$
means the exponential term containing the creation operators $a_{n}^{+}$ in
(8 )) taken in different positions (around a circle of length $L$) can be
expanded in terms of Macdonald polynomials attached to all possible Young
tableaux.\ The coefficients of these polynomials are, for each Young tableau 
$\lambda $, a combination of the creation operators $a_{n}^{+}\;$which we
call $a_{\lambda }^{+}\left( q,t\right) $.\ Then, the states $a_{\lambda
}^{+}\left( q,t\right) \mid \Omega _{N}>,$ where $\mid \Omega _{N}>$ is the
vacuum state for $N$ particles, define an orthonormal basis which is a
momentum space representation for the ''composite particles''.

\bigskip

In section 4, we propose a position space representation for the ''composite
particles''. The normal order product $:V\left( z_{1}\right) ...V\left(
z_{N}\right) :\;$which consists in writing all creation operators (including 
$Q)$ at the left of all annihilation operators (including $a_{0})$ provide a
position space representation for a set of different particles in different
positions. Here, we want to define a ''composite particles'' in a
complexified position $z$ with all its constituant particles in the same
position $z$; moreover, we wish to introduce a duality relation between the
quasihole and the ''composite particles'' (see Ref.$\left[ 10\right] )$. To
achieve this property, we define shifted positions $q^{i}z$ for the $N$
quasiparticles and quasiholes; then, if we choose $t=q^{N},$ we obtain the
property that the vertex operator for the ''composite particles'' is the
dual ($q\leftrightarrow t$ and change of sign$)$ of the vertex operator for
the quasihole. Now, the characteristic feature which tells that the
quasiparticles organize themselves into $p$ ''composite particles'' is that
these ones develop a discrete $Z_{p}$ symmetry. This symmetry can be
achieved using a formalism developped by Uglov$^{\left[ 20\right] }$ which
consists in taking the limit $q\rightarrow \exp \left( 2i\pi /p\right) $.

In the special case where $m=2n\;$ is even (in that case the ''composite
particles'' are bosons) and where there is a unique quasiparticle or
quasihole ($r=1)$ in the uncomplete ''composite particle'',$\;$we obtain for 
$p=1,$ the Laughlin ground state wave function 
\begin{equation}
<V_{N_{\pm }}\left( z_{1}\right) V_{N_{\pm }}\left( z_{2}\right) >=\left(
z_{1}-z_{2}\right) ^{2n\pm 1}  \label{11}
\end{equation}
corresponding to a filling factor $\nu =\frac{1}{2n\pm 1}.$ More generally,
for any $p,$ we obtain for the ground state wave function the product of a $%
Z_{p}$ invariant bosonic wave function times a fermionic wave function with
power $\pm 1$ depending whether we have one extra quasiparticle or quasihole 
\begin{equation}
<V_{N_{\pm }}\left( z_{1}\right) V_{N_{\pm }}\left( z_{2}\right) >=\left(
z_{1}^{p}-z_{2}^{p}\right) ^{2n}\;\left( z_{1}-z_{2}\right) ^{\pm 1}
\label{12}
\end{equation}
and corresponding to a filling factor $\nu =\frac{p}{2np\pm 1}.$

\bigskip

These results seem to be appropriate to describe a physic which is sensible
to the number of ''composite particles'' rather than to the number of the
individual particles which constitute them.

\bigskip

\bigskip

\section{Young tableaux and ''composite particles''}

\bigskip

\bigskip

Let us remind the reader that the $N$ particles are distributed over $p$
''composite particles'' containing $m$ particles each, and one extra
uncomplete ''composite particle'' containing $r$ particles ($0<r\leq m$) so
that $N=pm+r.$ The states of the ($p+1)\;$''composite particles'' are
labelled from top equal to $1$ to bottom equal to $d$; for each ''composite
particle'' we label the states from top equal to $1\;$to bottom equal to $%
\delta $, so that $\sum_{k=1}^{p+1}\;\delta _{k}=d.\;$To each state $i$ we
attach a momentum which is simply $(i-1)$ and which runs from $0$ to ($d-1)$%
. We label the particles from top to bottom and from left to right (on a
same state) by an integer $j$ which runs from $1$ to $N.$

In this section, we wish to show that for a given set of integers $l$ and $m$%
, and for a given number of states $d,$ there exists a one-to-one
correspondance between the possible configurations of $N$ particles into
''composite particles'' and the Young tableaux with at most $N\;$rows and
with the length of the first row $\lambda _{1}\leq d-pl-1$.

\bigskip The ground state (fig.2) is the unique configuration with $%
d=pl+1;\; $if $d\geq pl+1,$ the ground state is characterized by the fact
that the momentum of the particle $j$ is equal to $k_{j}^{0}=lE\left( \frac{%
j-1}{m}\right) .$ To the ground state, we naturally associate the empty
Young tableau where the length of the rows are all nul 
\begin{equation}
\lambda _{j}=0\;\;\;\;\;j=1,...,N  \label{13}
\end{equation}
Now, to a given configuration of particles in ''composite particles'' we
associate a Young tableau in the following way: to any particle $j$ of the
configuration, we associate a row of the Young tableau with length $\lambda
_{N-j+1}$ which is the difference between the momentum of $j$ in the given
configuration and the momentum of $j$ in the ground state configuration. If
there are $r_{i}$ particles on the same state $i,$ it corresponds to $r_{i}$
rows with the same length in the Young tableau. The Young tableau may be
partitionned from the bottom into $p$ blocks of $m$ rows each corresponding
to each ''composite particle'', and an extra block of $r$ rows at the top of
the tableau corresponding to the uncomplete ''composite particle''. It may
be noted that the length of the upper row of a given block is equal to the
length of the lower row of the next block up (if the corresponding state is
non empty); we also note that the length of the first row of the Young
tableau is $\lambda _{1}=k_{N}-pl$ where $k_{N}$ is the momentum of the $%
N^{th}$ particle. We give an illustration of the correspondance between
''composite particles'' and Young tableaux in fig.3. Of course, this
correspondance is crucially dependant of the values of the integers $l$ and $%
m$; in fig.4, we show how a given Young tableau corresponds to different
configurations for different choices of $\left( l,m\right) .$

\bigskip

\bigskip

\bigskip

\bigskip

We now introduce the shifted momenta as a consequence of the so-called Bethe
Ansatz which applies to the Calogero-Sutherland and Ruijsenaars-Schneider
models. The Bethe Ansatz equations are of the type 
\begin{equation}
e^{2i\pi \widetilde{k_{r}}}=\prod_{s\neq r}\;S\left( \widetilde{k_{s}}-%
\widetilde{k_{r}}\right)  \label{14}
\end{equation}
If we take the logarithm on both sides of (14), we get 
\begin{equation}
\widetilde{k_{r}}=k_{r}+\sum_{s\neq r}\Theta \left( \widetilde{k_{s}}-%
\widetilde{k_{r}}\right)  \label{15}
\end{equation}
where the $k_{r}$'s are integers which may be used to label the eigenstates
and where the phase shifts $\Theta \left( \widetilde{k}\right) $ are 
\begin{equation}
\Theta \left( \widetilde{k}\right) =\frac{1}{2i\pi }\ln S\left( \widetilde{k}%
\right)  \label{16}
\end{equation}
In Calogero-Sutherland and Ruijsenaars-Schneider models, we have 
\begin{eqnarray}
\Theta \left( \widetilde{k_{s}}-\widetilde{k_{r}}\right) &=&\frac{l}{m}%
\;if\;s<r  \label{17} \\
&=&0\;\;if\;s\geq r  \nonumber
\end{eqnarray}
Consequently, to a particle $j\;$inside a given ''composite particle'' we
attach a shifted momentum which is

\begin{equation}
\widetilde{k_{j}}=k_{j}+q\;\frac{l}{m}  \label{18}
\end{equation}
where $k_{j}$ is the momentum which labels the states and $q=0,1,...,m-1\;$
is the number of particles which are on the left (same state) and above $j,$
inside the corresponding ''composite particle''. If we introduce the
corresponding momentum $k_{j}^{0}$ of the particle in the ground state, we
obtain 
\begin{equation}
\widetilde{k_{j}}=\lambda _{N-j+1}+(mE\left( \frac{j-1}{m}\right) +q)\;\frac{%
l}{m}=\lambda _{N-j+1}+(j-1)\frac{l}{m}  \label{19}
\end{equation}
In order to make the correspondance with the usual momenta in the model of
Calogero-Sutherland, we still have to perform a global shift over all
momenta equal to $-\frac{N-1}{2}\frac{l}{m}$ because the fermi sea (ground
state for the fermions $l=m=1)$ used to have momenta which runs from $-\frac{%
N-1}{2}$ to $+\frac{N-1}{2}$ and not from $0$ to $N-1.$ We obtain the
characteristic expression for the Calogero-Sutherland momenta 
\begin{equation}
\widetilde{k_{j}}=\lambda _{N-j+1}+\left( j-\frac{N+1}{2}\right) \beta
\label{20}
\end{equation}
where the Calogero-Sutherland coupling constant is $\beta =\frac{l}{m}.$

\bigskip

The same organisation applies identically to the relativistic model of
Ruijsenaars-Schneider with the single difference that the momenta are $%
sh\left( \widetilde{k_{j}}\right) $ where $\widetilde{k_{j}}$ are the
shifted rapidities.

\bigskip

\section{The states for ''composite particles'' and vertex operators.}

\bigskip

We wish to use our complete knowledge of the eigenstates of the
Calogero-Sutherland and Ruijsenaars-Schneider Hamiltonian to construct the
space of states for the ''composite particles''. The technique used here is
the description of the eigenstates from the bosonization procedure and from
the corresponding vertex operators.

We first remind the reader the description of the eigenstates in the model
of Calogero-Sutherland.\ We introduce the complex variables 
\begin{equation}
z_{i}=\exp \left( \frac{2i\pi }{L}x_{i}\right)  \label{21}
\end{equation}
where $x_{i}$ is the circular coordinates of the particle $i$ along the
circle of length $L.\;$The ground state is represented by the function 
\begin{equation}
\Delta \left( z\right) =\prod_{i<j}\left( z_{i}-z_{j}\right) ^{\beta }
\label{22}
\end{equation}
where $\beta $ is the coupling constant in the Hamiltonian. Using the
technique of bosonisation, the above function can be represented in the
following way: we introduce the vertex operators 
\begin{equation}
V\left( z\right) =e^{\sqrt{\beta }Q}\exp \left( \sqrt{\beta }\sum_{n>0}\frac{%
a_{n}^{+}}{n}z^{n}\right) \;z^{\sqrt{\beta }a_{0}}\exp \left( -\sqrt{\beta }%
\sum_{n>0}\frac{a_{n}}{n}z^{-n}\right)  \label{23}
\end{equation}
where the creation operators $a_{n}^{+}\;$and the annihilation operators $%
a_{n}$ satisfy the commutation relations 
\begin{equation}
\left[ a_{n},a_{m}^{+}\right] =n\;\delta _{n,m}  \label{24}
\end{equation}
and where the operators $a_{0}$ and $Q$ satisfy 
\begin{equation}
\left[ a_{0},Q\right] =1  \label{25}
\end{equation}
all other commutators being nul.

We define the normal product 
\begin{equation}
:V\left( z_{1}\right) ...V\left( z_{N}\right) :  \label{26}
\end{equation}
as the product of the exponential operators where all exponentials for
creation operators (including $Q)\;$are written at the left of all
exponentials for annihilation operators (including $a_{0})$. Clearly, we
have 
\begin{equation}
V\left( z_{1}\right) ...V\left( z_{N}\right) =\prod_{i<j}\left(
z_{i}-z_{j}\right) ^{\beta }\;:V\left( z_{1}\right) ...V\left( z_{N}\right) :
\label{27}
\end{equation}
We define a vacuum state $\mid \Omega >$ such that the action of all the
annihilation operators upon $\mid \Omega >\;$gives zero 
\begin{equation}
a_{n}\mid \Omega >=0,\;\;n\geq 0  \label{28}
\end{equation}
Now, the action of the operator $\exp \left( \sqrt{\beta }Q\right) $ insures
the charge conservation in the vacuum expectation values and defines a
vacuum state for $N$ particles as 
\begin{equation}
e^{N\sqrt{\beta }Q}\mid \Omega >=\mid \Omega \left( N\sqrt{\beta }\right) >
\label{29}
\end{equation}

As a result of the above definitions, the ground state wave function is
found to be 
\begin{equation}
<\Omega \left( N\sqrt{\beta }\right) \mid V\left( z_{1}\right) ...V\left(
z_{N}\right) \mid \Omega >=\prod_{i<j}\left( z_{i}-z_{j}\right) ^{\beta }
\label{30}
\end{equation}
and we naturally associate the vacuum state $\mid \Omega \left( N\sqrt{\beta 
}\right) >$ ,with $\beta =\frac{l}{m},$ to the configuration of ''composite
particles'' which describes the ground state (fig.2).

The wave functions for the excited states of the Calogero-Sutherland
Hamiltonian are 
\begin{equation}
J_{\lambda }\left( z_{i},\beta \right) \;\prod_{i<j}\left(
z_{i}-z_{j}\right) ^{\beta }  \label{31}
\end{equation}
where $J_{\lambda }\left( z_{i},\beta \right) $ is the symmetric,
homogeneous Jack polynomial$^{\left[ 19\right] }$ attached to the Young
tableaux $\lambda $ which describes the momenta of the corresponding excited
state. Now, in the bosonisation procedure 
\begin{equation}
:V\left( z_{1}\right) ..V\left( z_{N}\right) :\mid \Omega >=\exp \left[ 
\sqrt{\beta }\sum_{n>0}\frac{a_{n}^{+}}{n}\left(
\sum_{i=1}^{N}z_{i}^{n}\right) \right] \;\mid \Omega \left( N\sqrt{\beta }%
\right) >  \label{32}
\end{equation}
can be expanded in terms of Jack polynomials; this expansion defines a set
of orthogonal states\ $a_{\lambda }^{+}\left( \beta \right) \mid \Omega
\left( N\sqrt{\beta }\right) >$ such that 
\begin{equation}
:V\left( z_{1}\right) ...V\left( z_{N}\right) :\mid \Omega >=\sum_{\lambda
}\;\sqrt{N_{\lambda }\left( \beta \right) }\;J_{\lambda }\left( z_{i},\beta
\right) \;a_{\lambda }^{+}\left( \beta \right) \;\mid \Omega \left( N\sqrt{%
\beta }\right) >  \label{33}
\end{equation}
where $N_{\lambda }\left( \beta \right) $ is a normalisation. As a
consequence, the wave functions for excited states (31) may be written as 
\begin{equation}
\frac{1}{_{\sqrt{N_{\lambda }\left( \beta \right) }}}<\Omega \left( N\sqrt{%
\beta }\right) \mid a_{\lambda }\left( \beta \right) \;V\left( z_{1}\right)
...V\left( z_{N}\right) \mid \Omega >  \label{34}
\end{equation}

\bigskip

From the facts that the configurations for ''composite particles'' are in a
one-to-one correspondance with the Young tableaux which describe the
eigenstates of the Calogero-Sutherland Hamiltonian, we may conclude that the
orthogonal basis $a_{\lambda }^{+}\left( \beta \right) \mid \Omega \left( N%
\sqrt{\beta }\right) >$ defines a representation of the configurations for
the ''composite particles''. This representation is a momentum space
representation; the adjoint representation in the complexified position
space is obtained from the primary fields $V\left( z\right) $ and their
products.\bigskip

\bigskip

The property that the states $\ a_{\lambda }^{+}\mid \Omega \left( N\sqrt{%
\beta }\right) >\;$form an orthonormal basis is non trivial.\ Another set of
symmetric, homogeneous polynomials which has this remarkable property is the
set of Macdonald's polynomials$^{\left[ 19\right] }\;M_{\lambda }\left(
z_{i},q,t\right) $. These polynomials depend of two parameters $q$ and $%
t=q^{\beta }$ and are generated from the vertex operators 
\begin{equation}
:V\left( z_{1}\right) ...V\left( z_{N}\right) :\mid \Omega >=\sum_{\lambda
}\;\sqrt{N_{\lambda }\left( q,t\right) }\;M_{\lambda }\left(
z_{i},q,t\right) \;a_{\lambda }^{+}\left( q,t\right) \mid \Omega \left( N%
\sqrt{\beta }\right) >  \label{35}
\end{equation}
where $N_{\lambda }\left( q,t\right) \;$is a normalisation.The vertex
operators in (35) are defined as 
\begin{equation}
V\left( z\right) =e^{\sqrt{\beta }Q}\exp \left( \sum_{n>0}\sqrt{\frac{1-t^{n}%
}{1-q^{n}}}\frac{a_{n}^{+}}{n}z^{n}\right) \;z^{\sqrt{\beta }a_{0}}\exp
\left( -\sum_{n>0}\sqrt{\frac{1-t^{n}}{1-q^{n}}}\frac{a_{n}}{n}z^{-n}\right)
\label{36}
\end{equation}
In (36), the commutation relations between creation and annihilation
operators are the same as in (24-25). In the case where $t$ and $q$ are
real, the wave function for the ground state may be found as 
\begin{equation}
<\Omega \left( N\sqrt{\beta }\right) \mid V\left( z_{1}\right) ...V\left(
z_{N}\right) \mid \Omega >=\prod_{i<j}\frac{\left( \frac{z_{j}}{z_{i}}%
;q\right) _{\infty }}{\left( t\frac{z_{j}}{z_{i}};q\right) _{\infty }}%
\prod_{j=1}^{N}z_{j}^{\left( N-j\right) \beta }  \label{37}
\end{equation}
where 
\begin{equation}
\left( x;q\right) _{n}=\prod_{i=0}^{n-1}(1-q^{i}x)  \label{38}
\end{equation}
Let us make the following remark: if $t=q^{k},\;k\in N_{+}$ and if $q$ is
real, then 
\begin{equation}
<\Omega \left( N\sqrt{\beta }\right) \mid V\left( z_{1}\right) ...V\left(
z_{N}\right) \mid \Omega >=\prod_{i<j}\prod_{r=0}^{k-1}\left(
z_{i}-q^{r}z_{j}\right)  \label{39}
\end{equation}
and especially if $q\rightarrow 1,$we get$\;\prod_{i<j}\left(
z_{i}-z_{j}\right) ^{k}.$\ 

When $t\;$and $q$ are real, the wave functions for the excited states are
obtained from 
\begin{eqnarray}
\frac{1}{\sqrt{N_{\lambda }\left( q,t\right) }} &<&\Omega \left( N\sqrt{%
\beta }\right) \mid a_{\lambda }\left( q,t\right) \;V\left( z_{1}\right)
...V\left( z_{N}\right) \mid \Omega >  \label{40} \\
&=&M_{\lambda }\left( z_{i},q,t\right) \;\prod_{i<j}\frac{\left( \frac{z_{j}%
}{z_{i}};q\right) _{\infty }}{\left( t\frac{zj}{z_{i}};q\right) _{\infty }}%
\prod_{j=1}^{N}z_{j}^{\left( N-j\right) \beta }  \nonumber
\end{eqnarray}

\bigskip

For \ this \ system \ again, \ we \ may \ conclude \ that \ the \
orthonormal \ basis$\;a_{\lambda }^{+}\left( q,t\right) \mid \Omega \left( N%
\sqrt{\beta }\right) >$ is a momentum space representation of the
configurations for the ''composite particles''. The primary fields $V\left(
z\right) $ and their products form an adjoint representation in the
complexified position space.

\bigskip

\bigskip

\section{Representation of ''composite particles'' in position space for $%
l=1 $.}

\bigskip

In this section, we wish to find a representation in position space for $%
N=pm+r\;\;(0<r\leq m)\;$ particles inside $p$ ''composite particles'' (+ one
uncomplete), the whole thing being defined at the complexified coordinate $z$%
. In order to construct this representation, we use the vertex operators
introduced in section 3. Then, we apply and extend our results to the case
where $N=pm\pm 1$ with $m$ even and interpret the theory as a possible
theory for the edge states of the quantum Hall effect at filling factor $\nu
=\frac{p}{pm\pm 1}$.

We follow in our construction a similar approach as the one proposed in Ref.$%
\left[ 10\right] $ by van Elburg and Schoutens. The quasihole with charge $%
-e/N>0$ and the electron with charge $e<0$ are defined as mutually dual
quasiparticles. More precisely, in the Calogero-Sutherland model, duality
means the exchange $\sqrt{\beta }\rightarrow -\frac{1}{\sqrt{\beta }}$ and
has for consequence on the wave functions of the excited states that the
Young tableaux $\lambda $ attached to the Jack polynomials are transformed
into $\widetilde{\lambda \text{ }}$ symmetric of $\lambda $ with respect to
its diagonal.

The authors of Ref.$\left[ 10\right] $ consider mainly the case where the
filling factor is $\nu =\frac{1}{m};$ in appendix A, they ''briefly describe
a quasiparticle formulation of the composite edge theories corresponding to
the filling fractions $\nu =\frac{p}{pm+1}$ of the Jain series''. They
consider that a convenient basis should be one ''which separates a single
charged mode from a set of $\left( p-1\right) $ neutral modes, the latter
being governed by an $\widehat{su\left( p\right) _{1}}\;$affine Kac-Moody
symmetry''. This program is achieved below (and extended to the case where
the filling factor is $\nu =\frac{p}{pm-1}$) by performing the limit
introduced by Uglov$^{\left[ 20\right] }:q\rightarrow \exp (2i\pi /p)$ in
the model of Ruijsenaars-Schneider.

The electron is a ''composite particles'' built from $p\;$bosonic
''composite particles'' each containing $m$ (even) quasiparticles + one
extra quasiparticle (for $\nu =\frac{p}{pm+1}$) or quasihole (for $\nu =%
\frac{p}{pm-1}$) as the uncomplete ''composite particle'' ($r=1);$ while the
electrons have a fermionic behaviour in the exchange of their positions, the
quasiparticles and the quasiholes exhibit in this exchange a behaviour
typical of a fractional statistic.

\bigskip\ 

We define the vertex operators for the quasihole $h$ and the quasiparticle $%
\pi $ as 
\begin{eqnarray}
V_{h}\left( z\right) &=&e^{-\frac{1}{\sqrt{\beta }}Q}\exp \left(
-\sum_{n>0}\varepsilon _{n}\chi _{n}^{\frac{1}{2}}\left( q,t\right) \;\frac{%
a_{n}^{+}}{n}\;z^{n}\right) .  \nonumber \\
&&\;.z^{-\frac{1}{\sqrt{\beta }}a_{0}}\exp \left( \sum_{n>0}\varepsilon
_{n}\chi _{n}^{\frac{1}{2}}\left( q,t\right) \frac{a_{n}}{n}\;z^{-n}\right) 
\eqnum{41}
\end{eqnarray}
\begin{eqnarray}
V_{\pi }\left( z\right) &=&e^{\frac{1}{\sqrt{\beta }}Q}\exp \left(
\sum_{n>0}\varepsilon _{n}\chi _{n}^{\frac{1}{2}}\left( q,t\right) \frac{%
a_{n}^{+}}{n}\;z^{n}\right) .  \nonumber \\
&&.\;z^{\frac{1}{\sqrt{\beta }}a_{0}}\exp \left( -\sum_{n>0}\varepsilon
_{n}\chi _{n}^{\frac{1}{2}}\left( q,t\right) \frac{a_{n}}{n}\;z^{-n}\right) 
\eqnum{42}
\end{eqnarray}
where $t=q^{\beta },$and 
\begin{equation}
\chi _{n}\left( q,t\right) =\left| \frac{q^{n/2}-q^{-n/2}}{t^{n/2}-t^{-n/2}}%
\right|  \eqnum{43}
\end{equation}
The symbol $\varepsilon _{n}$ defines the determination of the square roots
when $\left| q\right| =\left| t\right| =1$. We choose $\varepsilon _{n}=e^{i%
\frac{\pi }{2}\delta _{n}}$ \ \ \ where $\ \ \delta _{n}=0$ \ or $1$ \
depending whether $\left( q^{n/2}-q^{-n/2}\right) /\left(
t^{n/2}-t^{-n/2}\right) $ is $>0$ or $<0.$ The consequence of this choice is
that 
\begin{equation}
\left[ \varepsilon _{n}\chi _{n}^{\frac{1}{2}}\left( q,t\right) \right] ^{2}=%
\frac{q^{n/2}-q^{-n/2}}{t^{n/2}-t^{-n/2}}  \eqnum{44}
\end{equation}

This choice is done in order to avoid absolute values $\left| \;\;\right| $
and to preserve the analiticity in $q$ and $t$ when summing over $n$ in the
calculation of the product of vertex operators.

The duality property can be implemented in the theory of
Ruijsenaars-Schneider by shifting the variable $z\rightarrow q^{i}z$ and by
taking the product over $i$ as we show below.\ We define the vertex
operators for a set of $N_{+}=pm+r$ quasiparticles and for a set of $pm\;$%
quasiparticles and $r$ quasiholes; we define $N_{-}=pm-r>0$. By definition, 
\begin{equation}
V_{N_{+}}\left( z\right) =:V_{\pi }(q^{-\frac{N_{+}-1}{2}}z)V_{\pi }\left(
q^{-\frac{N_{+}-3}{2}}z\right) ...V_{\pi }\left( q^{\frac{N_{+}-1}{2}%
}z\right) \;:  \eqnum{45}
\end{equation}
\begin{equation}
V_{N_{-}}\left( z\right) =:V_{\pi }(q^{-\frac{N_{-}-1}{2}}z)...V_{\pi
}\left( q^{\frac{N_{+}-1}{2}}z\right) V_{h}\left( q^{\frac{N_{+}-1}{2}%
}z\right) ...V_{h}\left( q^{\frac{N_{-}+1}{2}}z\right) :  \eqnum{46}
\end{equation}
If we choose respectively $t=q^{N_{\pm }}$ that is $\beta =N\pm $, we get 
\begin{equation}
V_{N\pm }\left( z\right) =e^{\sqrt{\beta }Q}\exp \left(
\sum_{n>0}\varepsilon _{n}^{-1}\chi _{n}^{\frac{1}{2}}\left( t,q\right) 
\frac{a_{n}^{+}}{n}z^{n}\right) \;z^{\sqrt{\beta }a_{0}}\exp \left(
-\sum_{n>0}\varepsilon _{n}^{-1}\chi _{n}^{\frac{1}{2}}\left( t,q\right) 
\frac{a_{n}}{n}z^{-n}\right)  \eqnum{47}
\end{equation}
Clearly, the $N_{\pm }$ quasiparticles are described by a vertex operator
which is dual to the vertex operator for the quasihole. From (47), we obtain
the ground state wave functions 
\begin{equation}
<\Omega (2\sqrt{N_{\pm }})\mid V_{N_{\pm }}\left( z_{1}\right) V_{N_{\pm
}}\left( z_{2}\right) \mid \Omega >=\prod_{k=-\left( N_{\pm }-1\right)
/2}^{\left( N_{\pm }-1\right) /2}\left( z_{1}-q^{k}\;z_{2}\right)  \eqnum{48}
\end{equation}
where the sum over $k$ is step one, all $k^{\prime }s$ are integers if $N$
is odd and are half integers if $N$ is even. The duality property gives 
\begin{equation}
<\Omega \left( \sqrt{N_{\pm }}-\frac{1}{\sqrt{N_{\pm }}}\right) \mid
V_{N_{\pm }}\left( z_{1}\right) V_{h_{\pm }}\left( z_{2}\right) \mid \Omega
>=\left( z_{1}-z_{2}\right) ^{-1}  \eqnum{49}
\end{equation}
while 
\begin{equation}
<\Omega \left( \sqrt{N_{\pm }}+\frac{1}{\sqrt{N_{\pm }}}\right) \mid
V_{N_{\pm }}\left( z_{1}\right) V_{\pi _{\pm }}\left( z_{2}\right) \mid
\Omega >=\left( z_{1}-z_{2}\right)  \eqnum{50}
\end{equation}

If the quasiparticles and the quasiholes are not organized in ''composite
particles'' we may take the limit $q\rightarrow 1$ and we obtain 
\begin{equation}
V_{h_{\pm }}\left( z\right) =e^{-\frac{1}{\sqrt{N_{\pm }}}Q}\exp \left( -%
\frac{1}{\sqrt{N_{\pm }}}\sum_{n>0}\frac{a_{n}^{+}}{n}z^{n}\right) \;z^{-%
\frac{1}{\sqrt{N_{\pm }}}a_{0}}\exp \left( \frac{1}{\sqrt{N_{\pm }}}%
\sum_{n>0}\frac{a_{n}}{n}z^{-n}\right)  \eqnum{51}
\end{equation}
\begin{equation}
V_{\pi _{\pm }}\left( z\right) =e^{\frac{1}{\sqrt{N_{\pm }}}Q}\exp \left( 
\frac{1}{\sqrt{N_{\pm }}}\sum_{n>0}\frac{a_{n}^{+}}{n}z^{n}\right) \;z^{%
\frac{1}{\sqrt{N_{\pm }}}a_{0}}\exp \left( -\frac{1}{\sqrt{N_{\pm }}}%
\sum_{n>0}\frac{a_{n}}{n}z^{-n}\right)  \eqnum{52}
\end{equation}
\begin{equation}
V_{N_{\pm }}\left( z\right) =e^{\sqrt{N_{\pm }}Q}\exp \left( \sqrt{N_{\pm }}%
\sum_{n>0}\frac{a_{n}^{+}}{n}z^{n}\right) \;z^{\sqrt{N_{\pm }}a_{0}}\exp
\left( -\sqrt{N_{\pm }}\sum_{n>0}\frac{a_{n}}{n}z^{-n}\right)  \eqnum{53}
\end{equation}
By choosing $t=q^{N_{\pm }},$ the vertex operators for the quasiholes and
for the quasiparticles become $N_{\pm }$ dependant so that they know that
they belong to a set of $N_{+}$ quasiparticles or a set of $pm$
quasiparticles and $r$ quasiholes; in the limit $q\rightarrow 1,$ the $r$
quasiholes simply destroy $r$ quasiparticles. If $0<r\leq m,$ the vertex
operators $V_{N_{+}}\left( z\right) $ provide a description of the
''uncomplete composite particles'' alone, that is for $N_{+}\leq m$ or $p=0.$
If we assign to $V_{h_{\pm }}\left( z\right) $ a charge $+1$, the operator $%
V_{\pi _{\pm }}\left( z\right) $ has a charge $-1$ and the operator $%
V_{N_{\pm }}\left( z\right) $ has a charge $-N_{\pm }$.

We obtain the following operator products 
\begin{equation}
V_{\pi _{\pm }}\left( z_{1}\right) V_{\pi _{\pm }}\left( z_{2}\right)
=\left( z_{1}-z_{2}\right) ^{\frac{1}{N_{\pm }}}:V_{\pi _{\pm }}\left(
z_{1}\right) V_{\pi _{\pm }}\left( z_{2}\right) :  \eqnum{54}
\end{equation}
\begin{equation}
V_{N_{\pm }}\left( z_{1}\right) V_{N_{\pm }}\left( z_{2}\right) =\left(
z_{1}-z_{2}\right) ^{N_{\pm }}:V_{N_{\pm }}\left( z_{1}\right) V_{N_{\pm
}}\left( z_{2}\right) :  \eqnum{55}
\end{equation}
\begin{equation}
V_{N_{\pm }}\left( z_{1}\right) V_{\pi _{\pm }}\left( z_{2}\right) =\left(
z_{1}-z_{2}\right) :V_{N_{\pm }}\left( z_{1}\right) V_{\pi _{\pm }}\left(
z_{2}\right) :  \eqnum{56}
\end{equation}
and the ground state wave functions for the $N_{+}$ quasiparticles or the $%
pm $ quasiparticles and the $r$ quasiholes are 
\begin{equation}
<\Omega \left( 2\sqrt{N_{\pm }}\right) \mid V_{N_{\pm }}\left( z_{1}\right)
V_{N_{\pm }}\left( z_{2}\right) \mid \Omega >=\left( z_{1}-z_{2}\right)
^{N_{\pm }}  \eqnum{57}
\end{equation}
This wave function has a zero of order $N_{\pm }$ at $z_{1}=z_{2}\,%
\;(x_{1}=x_{2}$ on the circle) that is $N_{\pm }$ times the vanishing
property for one particle $(N_{\pm }=1).$

\bigskip

Now, in the general case where the$\;pm\;$quasiparticles organize themselves
in $p$ ''composite particles'' of $m$ quasiparticles each, we must implement
the fact that the $p$ ''composite particles'' have an extra symmetry $Z_{p}.$
This can be achieved using a formalism due to Uglov$^{\left[ 20\right] }$
and which consists in taking the limit $q\rightarrow \exp \left( 2i\pi
/p\right) $ in (41,42,47) together with $t=q^{N_{\pm }}.$ In this limit we
have in (48) a reorganization of the product over $k$ by grouping $m$ times
a product of $p$ quantities, times (or divided by) a product of $r$
quantities: 
\begin{equation}
<\Omega \left( 2\sqrt{N_{\pm }}\right) \mid V_{N_{\pm }}\left( z_{1}\right)
V_{N_{\pm }}\left( z_{2}\right) \mid \Omega >=\left( z_{1}^{p}+\left(
-\right) ^{N_{\pm }}z_{2}^{p}\right) ^{m}\left[ \prod_{k=-\frac{r-1}{2}}^{%
\frac{r-1}{2}}\left( z_{1}-\left( -\right) ^{m}q^{k}z_{2}\right) \right]
^{\pm 1}  \eqnum{58}
\end{equation}
This is the ground state wave function for a ''composite particles'' of $%
N_{+}$ quasiparticles or of $pm$ quasiparticles and $r$ quasiholes $\left(
N_{-}=pm-r\right) .$

\bigskip

\bigskip

We now decompose the vertex operators in $p$ different modes, one charged
mode and $\left( p-1\right) $ neutral modes. We write 
\begin{equation}
\varepsilon _{n}^{-1}\chi _{n}^{\frac{1}{2}}\left( q^{N_{\pm }},q\right)
\rightarrow \varepsilon _{k}^{-1}\sqrt{N_{\pm }}\;\;\;\;if\;n=kp,\;\;\;k\in Z
\eqnum{59}
\end{equation}
where $\varepsilon _{k}=e^{i\frac{\pi }{2}\delta _{k}}$ with $\delta _{k}=0$
or $1$ whether $\left( N-1\right) k$ is even or odd, and 
\begin{equation}
\varepsilon _{n}^{-1}\chi _{n}^{\frac{1}{2}}\left( q^{N_{\pm }},q\right)
=\varepsilon _{k,s}^{-1}\chi _{s}^{\frac{1}{2}}\left( q^{\pm r},q\right)
\;\;if\;n=kp+s,\;\ \ \ s=1,...,p-1\;\;\;\;k\in Z\;  \eqnum{60}
\end{equation}
where \ $\varepsilon _{k,s}=e^{i\frac{\pi }{2}\left[ (N_{\pm }-1)k+ms+\delta
_{\pm ,s}\right] }$ \ \ with $\;\delta _{\pm ,s}=0$ or $1\;\;$whether

$\left( q^{\pm rs/2}-q^{\mp rs/2}\right) /\left( q^{s/2}-q^{-s/2}\right) $
is $>0$ or $<0.$ We obtain the vertex operators $V_{\pi _{\pm }}\left(
z\right) $ and $V_{N_{\pm }}\left( z\right) $ as products of $p$ independant
vertex operators 
\begin{equation}
V_{N_{\pm }}\left( z\right) =:\overline{V_{\pm ,0}}\left( z^{p}\right)
\prod_{s=1}^{p-1}\overline{V_{\pm ,s}}\left( z^{p}\right) :  \eqnum{61}
\end{equation}
\begin{equation}
V_{\pi _{\pm }}\left( z\right) =:\overline{V_{\pm ,0}^{\prime }}\left(
z^{p}\right) \prod_{s=1}^{p-1}\overline{V_{\pm ,s}^{\prime }}\left(
z^{p}\right) :  \eqnum{62}
\end{equation}
with 
\begin{eqnarray}
\overline{V_{\pm ,0}}\left( z\right) &=&e^{\sqrt{\frac{N_{\pm }}{p}}%
\overline{Q}}\exp \left( \sqrt{\frac{N_{\pm }}{p}}\sum_{k>0}\varepsilon
_{k}^{-1}\frac{A_{k}^{+}}{k}z^{k}\right) .  \nonumber \\
&&.\;z^{\sqrt{\frac{N_{\pm }}{p}}A_{0}}\exp \left( -\sqrt{\frac{N_{\pm }}{p}}%
\sum_{k>0}\varepsilon _{k}^{-1}\frac{A_{k}}{k}z^{-k}\right)  \eqnum{63}
\end{eqnarray}
\begin{eqnarray}
\overline{V_{\pm ,s}}\left( z\right) &=&\exp \left( \frac{\chi _{s}^{\frac{1%
}{2}}\left( q^{\pm r},q\right) }{\sqrt{p}}\sum_{k\geq 0}\varepsilon
_{k,s}^{-1}\frac{A_{k+s/p}^{+}}{k+s/p}z^{k+s/p}\right) .  \nonumber \\
&&.\exp \left( -\frac{\chi _{s}^{\frac{1}{2}}\left( q^{\pm r},q\right) }{%
\sqrt{p}}\sum_{k\geq 0}\varepsilon _{k,s}^{-1}\frac{A_{k+s/p}}{k+s/p}%
z^{-\left( k+s/p\right) }\right)  \eqnum{64}
\end{eqnarray}
\begin{eqnarray}
\overline{V_{\pm ,0}^{\prime }}\left( z\right) &=&e^{\frac{1}{\sqrt{pN_{\pm }%
}}\overline{Q}}\exp \left( \frac{1}{\sqrt{pN_{\pm }}}\sum_{k\geq
0}\varepsilon _{k}\frac{A_{k}^{+}}{k}z^{k}\right) .  \nonumber \\
&&.\;z^{\frac{1}{\sqrt{pN_{\pm }}}A_{0}}\exp \left( -\frac{1}{\sqrt{pN_{\pm }%
}}\sum_{k\geq 0}\varepsilon _{k}\frac{A_{k}}{k}z^{-k}\right)  \eqnum{65}
\end{eqnarray}
\begin{eqnarray}
\overline{V_{\pm ,s}^{\prime }}\left( z\right) &=&\exp \left( \frac{\chi
_{s}^{\frac{1}{2}}\left( q,q^{\pm r}\right) }{\sqrt{p}}\sum_{k\geq
0}\varepsilon _{k,s}\frac{A_{k+s/p}^{+}}{k+s/p}z^{k+s/p}\right) .  \nonumber
\\
&&.\exp \left( -\frac{\chi _{s}^{\frac{1}{2}}\left( q,q^{\pm r}\right) }{%
\sqrt{p}}\sum_{k\geq 0}\varepsilon _{k,s}\frac{A_{k+s/p}}{k+s/p}z^{-\left(
k+s/p\right) }\right)  \eqnum{66}
\end{eqnarray}
In (63-66) we defined the creation and annihilation operators $%
A_{k+s/p},\;A_{k+s/p}^{+}$ and $\overline{Q}$ as 
\begin{equation}
\;a_{kp+s}=\sqrt{p}A_{k+s/p}  \eqnum{67}
\end{equation}
\begin{equation}
\;a_{kp+s}^{+}=\sqrt{p}A_{k+s/p}^{+}  \eqnum{68}
\end{equation}
\begin{equation}
Q=\frac{1}{\sqrt{p}}\overline{Q}  \eqnum{69}
\end{equation}
so that 
\begin{equation}
\left[ A_{k+s/p},A_{k^{\prime }+s^{\prime }/p}^{+}\right] =\left(
k+s/p\right) \;\delta _{k,k^{\prime }}\;\delta _{s,s^{\prime }}  \eqnum{70}
\end{equation}
\begin{equation}
\left[ A_{0},\overline{Q}\right] =1  \eqnum{71}
\end{equation}
It is clear that the operators $\overline{V_{\pm ,s}}\left( z\right) $ and $%
\overline{V_{\pm ,s^{\prime }}}\left( z^{\prime }\right) $ (as well as $%
\overline{V_{\pm ,s}^{\prime }}\left( z\right) $ and $\overline{V_{\pm
,s^{\prime }}^{\prime }}\left( z^{\prime }\right) $) commute for $s\neq
s^{\prime }.$ The operator $\overline{V_{\pm ,0}}\left( z\right) $ has a
charge $+N_{\pm }$ if the operator $\overline{V_{\pm ,0}^{\prime }}\left(
z\right) $ has a charge $+1$. The remaining $\left( p-1\right) $ operators $%
\overline{V_{\pm ,s}}\left( z\right) $ (or $\overline{V_{\pm ,s}^{\prime }}%
\left( z\right) $) are neutral. It is interesting to note that $V_{N_{\pm
}}\left( z\right) $ and $V_{\pi _{\pm }}\left( z\right) $ in (61,62) are
really functions of $z^{p};$ this will be commented later on$.$

\bigskip From (63,65), we successively obtain for the charged mode 
\begin{equation}
\overline{V_{\pm ,0}}\left( z_{1}\right) \overline{V_{\pm ,0}}\left(
z_{2}\right) =\left( z_{1}+\left( -\right) ^{N_{\pm }}z_{2}\right) ^{\frac{%
N_{\pm }}{p}}\;:\overline{V_{\pm ,0}}\left( z_{1}\right) \overline{V_{\pm ,0}%
}\left( z_{2}\right) :  \eqnum{72}
\end{equation}
\begin{equation}
\overline{V_{\pm ,0}^{\prime }}\left( z_{1}\right) \overline{V_{\pm
,0}^{\prime }}\left( z_{2}\right) =\left( z_{1}+\left( -\right) ^{N_{\pm
}}z_{2}\right) ^{\frac{1}{pN_{\pm }}}\;:\overline{V_{\pm ,0}^{\prime }}%
\left( z_{1}\right) \overline{V_{\pm ,0}^{\prime }}\left( z_{2}\right) : 
\eqnum{73}
\end{equation}
\begin{equation}
\overline{V_{\pm ,0}}\left( z_{1}\right) \overline{V_{\pm ,0}^{\prime }}%
\left( z_{2}\right) =\left( z_{1}-z_{2}\right) ^{\frac{1}{p}}\;:\overline{%
V_{\pm ,0}}\left( z_{1}\right) \overline{V_{\pm ,0}^{\prime }}\left(
z_{2}\right) :  \eqnum{74}
\end{equation}
Similarly, using the relation 
\begin{equation}
\sum_{k\geq 0}\frac{x^{k+\frac{s}{p}}}{k+\frac{s}{p}}=-%
\sum_{u=0}^{p-1}q^{-us}\ln \left[ 1-q^{u}x^{1/p}\right] ,\;\;\;s=0,...,p-1\;%
\;\;  \eqnum{75}
\end{equation}
where $q=\exp (2i\pi /p),$ we obtain for all the neutral modes \bigskip 
\begin{eqnarray}
W_{\pm }\left( z\right) &=&\prod_{s=1}^{p-1}\overline{V_{\pm ,s}}\left(
z\right)  \nonumber \\
W_{\pm }^{\prime }\left( z\right) &=&\prod_{s=1}^{p-1}\overline{V_{\pm
,s}^{\prime }}\left( z\right)  \eqnum{76}
\end{eqnarray}
the relations

\begin{eqnarray}
W_{\pm }\left( z_{1}\right) W_{\pm }\left( z_{2}\right) &=&\left(
z_{1}+\left( -\right) ^{N_{\pm }}z_{2}\right) ^{\mp \frac{r}{p}}\left[
\prod_{k=-\frac{r-1}{2}}^{\frac{r-1}{2}}\left( z_{1}^{1/p}-\left( -\right)
^{m}q^{k}z_{2}^{1/p}\right) \right] ^{\pm 1}.  \nonumber \\
. &:&W_{\pm }\left( z_{1}\right) W_{\pm }\left( z_{2}\right) :  \eqnum{77}
\end{eqnarray}
\begin{eqnarray}
W_{\pm }^{\prime }\left( z_{1}\right) W_{\pm }^{\prime }\left( z_{2}\right)
&=&\left( z_{1}+\left( -\right) ^{N_{\pm }}z_{2}\right) ^{\mp \frac{1}{pr}%
}\prod_{k=-\frac{r-1}{2}}^{\frac{r-1}{2}}\prod_{u=0}^{p-1}\left[
z_{1}^{1/p}-\left( -\right) ^{m}q^{k+u}z_{2}^{1/p}\right] ^{\pm \rho
_{r}\left( u\right) }.  \nonumber \\
. &:&W_{\pm }^{\prime }\left( z_{1}\right) W_{\pm }^{\prime }\left(
z_{2}\right) :  \eqnum{78}
\end{eqnarray}
where 
\begin{equation}
\ \rho _{r}(u)=\frac{1}{p}\sum_{s=0}^{p-1}q^{-us}\frac{\sin ^{2}\left( \frac{%
\pi s}{p}\right) }{\sin ^{2}\left( \frac{\pi rs}{p}\right) }  \eqnum{79}
\end{equation}
and 
\begin{equation}
W_{\pm }\left( z_{1}\right) W_{\pm }^{\prime }\left( z_{2}\right) =\frac{%
\left[ z_{1}^{1/p}-\;z_{2}^{1/p}\right] }{\left( z_{1}-z_{2}\right) ^{1/p}}%
:W_{\pm }\left( z_{1}\right) W_{\pm }^{\prime }\left( z_{2}\right) : 
\eqnum{80}
\end{equation}
$\ \ \ \ \ \ \ \ \ \ \ \ \ \ \ \ \ \ \ \ \ \ \ \ \ \ \ \ \ \ \ \ \ \ \ \ \ \
\ \ \ \ \ \ \ \ \ \ \ $

\bigskip

\bigskip

We now specify the special case where $N_{\pm }=pm\pm 1$ ($m$ even) and
justify the application to the fractional quantum Hall effect with filling
factor $\nu =p/(pm\pm 1).\;$Hopefully, the above results simplify greatly at 
$r=1$ since $\rho _{1}\left( u\right) =\delta _{u,0.}\;$ The vertex operator
for the electron is defined as 
\begin{equation}
V_{N_{\pm }}\left( z\right) =:e^{\sqrt{\frac{pm\pm 1}{p}}\phi _{0}\left(
z^{p}\right) +\frac{\varepsilon _{\pm }^{-1}}{\sqrt{p}}\sum_{s=1}^{p-1}\phi
_{s}\left( z^{p}\right) }:  \eqnum{81}
\end{equation}
The vertex operator for the quasiparticle is defined as 
\begin{equation}
V_{\pi _{\pm }}\left( z\right) =:e^{\sqrt{\frac{1}{p\left( pm\pm 1\right) }}%
\phi _{0}\left( z^{p}\right) +\frac{\varepsilon _{\pm }}{\sqrt{p}}%
\sum_{s=1}^{p-1}\phi _{s}\left( z^{p}\right) }:  \eqnum{82}
\end{equation}
The vertex operator for the quasihole is defined as 
\begin{equation}
V_{h_{\pm }}\left( z\right) =:e^{-\sqrt{\frac{1}{p\left( pm\pm 1\right) }}%
\phi _{0}\left( z^{p}\right) -\frac{\varepsilon _{\pm }}{\sqrt{p}}%
\sum_{s=1}^{p-1}\phi _{s}\left( z^{p}\right) }:  \eqnum{83}
\end{equation}
where $\varepsilon _{\pm }=e^{i\frac{\pi }{2}\delta _{\pm }}$ with $\delta
_{+}=0$ and $\delta _{-}=1.$ In (81-83), $\phi _{0}\left( z\right) =\phi
_{0}^{+}\left( z\right) +\phi _{0}^{-}\left( z\right) \;$with 
\begin{eqnarray}
\phi _{0}^{+}\left( z\right) &=&\overline{Q}+\sum_{k>0}\frac{A_{k}^{+}}{k}%
z^{k}  \nonumber \\
\phi _{0}^{-}\left( z\right) &=&A_{0}\;\ln z-\sum_{k>0}\frac{A_{k}}{k}z^{-k}
\eqnum{84}
\end{eqnarray}
so that 
\begin{equation}
\left[ \phi _{0}^{-}\left( z_{1}\right) ,\phi _{0}^{+}\left( z_{2}\right) %
\right] =\ln \left( z_{1}-z_{2}\right)  \eqnum{85}
\end{equation}
Similarly, $\phi _{s}\left( z\right) =\phi _{s}^{+}\left( z\right) +\phi
_{s}^{-}\left( z\right) $ with 
\begin{eqnarray}
\phi _{s}^{+}\left( z\right) &=&\sum_{k\geq 0}\frac{A_{k+s/p}}{k+s/p}%
z^{\left( k+s/p\right) },\;\;\;s=1,...,p-1  \nonumber \\
\phi _{s}^{-}\left( z\right) &=&-\sum_{k\geq 0}\frac{A_{k+s/p}}{k+s/p}%
z^{-\left( k+s/p\right) },\;s=1,...,p-1  \eqnum{86}
\end{eqnarray}
so that 
\begin{equation}
\left[ \phi _{s}^{-}\left( z_{1}\right) ,\phi _{s^{\prime }}^{+}\left(
z_{2}\right) \right] =\delta _{s,s^{\prime }}\;\sum_{u=0}^{p-1}q^{-us}\ln %
\left[ 1-q^{u}\left( \frac{z_{2}}{z_{1}}\right) ^{1/p}\right]  \eqnum{87}
\end{equation}
and consequently 
\begin{equation}
\left[ \sum_{s=1}^{p-1}\phi _{s}^{-}\left( z_{1}\right) ,\sum_{s^{\prime
}=1}^{p-1}\phi _{s^{\prime }}^{+}\left( z_{2}\right) \right] =\ln \left[ 
\frac{\left[ z_{1}^{1/p}-z_{2}^{1/p}\right] ^{p}}{\left( z_{1}-z_{2}\right) }%
\right]  \eqnum{88}
\end{equation}

\bigskip

We obtain the ground state wave functions for the electrons, the
quasiparticles and the quasiholes as 
\begin{equation}
<\Omega \left( 2\sqrt{N_{\pm }}\right) \mid V_{N\pm }\left( z_{1}\right)
V_{N\pm }\left( z_{2}\right) \mid \Omega >=\left( z_{1}^{p}-z_{2}^{p}\right)
^{m}\;\left( z_{1}-z_{2}\right) ^{\pm 1}  \eqnum{89}
\end{equation}
We note that at $p=1,$we obtain the so-called Laughlin wave function. 
\begin{equation}
<\Omega \left( \frac{2}{\sqrt{N_{\pm }}}\right) \mid V_{\pi \pm }\left(
z_{1}\right) V_{\pi \pm }\left( z_{2}\right) \mid \Omega >=\left(
z_{1}^{p}-z_{2}^{p}\right) ^{\frac{\mp m}{pm\pm 1}}\;\left(
z_{1}-z_{2}\right) ^{\pm 1}  \eqnum{90}
\end{equation}
\begin{equation}
<\Omega \left( \sqrt{N_{\pm }}+\frac{1}{\sqrt{N_{\pm }}}\right) \mid V_{N\pm
}\left( z_{1}\right) V_{\pi \pm }\left( z_{2}\right) \mid \Omega >=\left(
z_{1}-z_{2}\right)  \eqnum{91}
\end{equation}
\begin{equation}
<\Omega \left( \sqrt{N_{\pm }}-\frac{1}{\sqrt{N_{\pm }}}\right) \mid V_{N\pm
}\left( z_{1}\right) V_{h\pm }\left( z_{2}\right) \mid \Omega >=\left(
z_{1}-z_{2}\right) ^{-1}  \eqnum{92}
\end{equation}

\bigskip

\bigskip

The electron wave function is antisymmetric in the exchange $%
z_{1}\leftrightarrow z_{2}$ while the quasiparticle and the quasihole behave
in this exchange according to the fractional statistic described by the
exchange angle 
\begin{equation}
\frac{\theta }{\pi }=\pm 1-\frac{\pm m}{pm\pm 1}  \eqnum{93}
\end{equation}

\bigskip If we define 
\begin{equation}
m^{\prime }=\frac{\mp m}{pm\pm 1}  \eqnum{94}
\end{equation}
we have the relation 
\begin{equation}
pm^{\prime }\pm 1=\frac{1}{pm\pm 1}  \eqnum{95}
\end{equation}
which is another manifestation of the duality between the electron and the
quasihole. If the charge of the electron is $e$, then the charge of the
quasiparticle is $e^{\ast }=\frac{e}{N_{\pm }}=\frac{e}{pm\pm 1}=\pm e\left(
1-m\nu \right) $ where we introduce the filling ratio $\nu =\frac{p}{N_{\pm }%
}=\frac{p}{pm\pm 1}.$ The $U\left( 1\right) $ charge comes from the field $%
\phi _{0}\left( z\right) $ while the $\left( p-1\right) $ remaining modes $%
\phi _{s}\left( z\right) $ are neutral.\ The charged sector is described by
a Calogero-Sutherland model with $\beta =\frac{1}{\nu };$ the neutral
sectors are described by Calogero-Sutherland models with $\beta =\frac{1}{%
\pm p}$ and without the zero modes (the terms $\overline{Q\text{ }}$ and $%
A_{0}).$

\bigskip Let us mention here that we concentrated on the ground state wave
function for two set of ''composite particles'', but the expansion of the
products of vertex operators may provide the wave functions for the various
excited states and for several set of ''composite particles'' in different
complexified coordinates $z_{i}$.

\bigskip

\bigskip

Let us close this section by trying to understand the relation between the
fields $\phi _{s}\left( z\right) ,\;s=0,...,p-1$ and the $p$ fields usually
introduced in the Luttinger liquid in order to describe the edge states of
the quantum Hall effect. According to Wen$^{\left[ 24\right] }$ 
\begin{equation}
S_{edge}=\frac{1}{4\pi }\int dtdx\sum_{I,J=1}^{p}\;\left[ K_{IJ}\;\partial
_{t}\widetilde{\varphi _{I}}\;\partial _{x}\widetilde{\varphi _{J}}%
-V_{IJ}\;\partial _{x}\widetilde{\varphi _{I}}\;\partial _{x}\widetilde{%
\varphi _{J}}\right]  \eqnum{96}
\end{equation}
where $V$ is a positive definite matrix which describes the velocity of the
edge excitations, and where the $p\times p\;$matrix\ $K_{IJ}=m+\delta _{I,J}$%
.\ The cinetic part of the Lagrangian disappear in the Hamiltonian 
\begin{equation}
H_{edge}=\frac{1}{4\pi }\int dtdx\sum_{I,J=1}^{p}\;V_{IJ}\;\partial _{x}%
\widetilde{\varphi _{I}}\;\partial _{x}\widetilde{\varphi _{J}}  \eqnum{97}
\end{equation}
However, the cinetic part of the Lagrangian is responsible for the canonical
quantization of the system. The matrix $K_{IJ}\;$is cyclic with eigenvalues $%
\lambda _{0}=pm+1$ and $\lambda _{s}=1,\;s=1,...,p-1.$ We may diagonalize
the cinetic part by Fourier transforming it 
\begin{equation}
\widetilde{\varphi _{I}}=\frac{1}{p}\sum_{s=0}^{p-1}q^{-s\;I}\;\varphi _{s} 
\eqnum{98}
\end{equation}
with $q=\exp (2i\pi /p).$\ The cinetic part of the Lagrangian becomes 
\begin{equation}
L=\frac{pm+1}{p}\partial _{t}\varphi _{0}\partial _{x}\varphi _{0}+\frac{1}{p%
}\sum_{s=1}^{p-1}\partial _{t}\varphi _{s}\partial _{x}\varphi _{p-s} 
\eqnum{99}
\end{equation}
so that the conjugate momenta are 
\begin{eqnarray}
\pi _{0} &=&\frac{pm+1}{p}\partial _{x}\varphi _{0}  \nonumber \\
\pi _{s} &=&\frac{1}{p}\partial _{x}\varphi _{p-s}  \eqnum{100}
\end{eqnarray}
The equal time commutators 
\begin{equation}
\left[ \varphi _{s}\left( x,0\right) ,\pi _{s^{\prime }}\left( y,0\right) %
\right] =i\;\delta _{s,s^{\prime }}\;\delta \left( x-y\right)  \eqnum{101}
\end{equation}
give 
\begin{eqnarray}
\left[ \varphi _{0}\left( x,0\right) ,\varphi _{0}\left( y,0\right) \right]
&=&-\frac{i}{2}\;\;\frac{p}{pm+1}\;\;\epsilon \left( x-y\right)  \nonumber \\
\left[ \varphi _{s}\left( x,0\right) ,\varphi _{s^{\prime }}\left(
y,0\right) \right] &=&-\frac{i}{2}\;\;\delta _{s+s^{\prime
},0}\;\;p\;\;\epsilon \left( x-y\right)  \eqnum{102}
\end{eqnarray}
or equivalently 
\begin{equation}
\left[ \psi _{s}\left( x,0\right) ,\psi _{s^{\prime }}\left( y,0\right) %
\right] =i\pi \;\delta _{s+s^{\prime },0}\;\;\epsilon \left( x-y\right) 
\eqnum{103}
\end{equation}
with 
\begin{eqnarray}
\psi _{0}\left( x,t\right) &=&i\sqrt{2\pi }\;\sqrt{\frac{pm+1}{p}}\;\varphi
_{0}\left( x,t\right)  \nonumber \\
\psi _{s}\left( x,t\right) &=&i\sqrt{2\pi }\;\sqrt{\frac{1}{p}}\;\;\varphi
_{s}\left( x,t\right) ,\;\;\;\;s=1,...,p-1  \eqnum{104}
\end{eqnarray}
If we split $\psi _{s}\left( x,0\right) $ into annihilation and creation
parts and the distribution $\epsilon \left( x-y\right) $ accordingly, we get 
\begin{equation}
\left[ \psi _{s}^{-}\left( x,0\right) ,\psi _{s^{\prime }}^{+}\left(
y,0\right) \right] =\delta _{s+s^{\prime },0}\;\ln \left( e^{2i\pi \frac{x}{L%
}}-e^{2i\pi \frac{y}{L}}\right)  \eqnum{105}
\end{equation}
where $L$ is a regulator and $-L/2\leq x-y\leq L/2.$

For $s=s^{\prime }=0,$ the commutator (105) is the same than the commutator
(85). Consequently, the fields $\phi _{0}\left( z\right) $ and $\psi
_{s}\left( x,0\right) $ are unitarily equivalent. The main difference
between Wen's theory and our results is that the fields $\psi _{s}\left(
x,0\right) $ are equally charged (as can be seen from (105)) for $%
s=0,...,p-1,$ while our fields $\phi _{s}\left( z\right) $ are neutral for $%
s=1,...,p-1.$ One way of recovering one charged field and $p-1$ neutral
fields from Wen's fields $\psi _{s}\left( x,0\right) $ should be to
distinguish in their Fourier decomposition and in the decomposition of 
\begin{equation}
\epsilon \left( x-y\right) =\frac{1}{i\pi }\left[ \sum_{n\neq 0}\frac{1}{n}%
e^{2i\pi n\frac{x-y}{L}}\;+2i\pi \frac{x-y}{L}\right] \;  \eqnum{106}
\end{equation}
the various modes $n=kp$ and $n=kp+s,\;s=1,...,p-1$ but this is exactly what
we have done.

\bigskip

In Ref.$\left[ 21-22\right] $ the authors introduce an extra transformation $%
z\rightarrow z^{1/p}$ which is suggested by the fact that the vertex
operators $V_{N_{\pm }}\left( z\right) ,$ $V_{\pi _{\pm }}\left( z\right) \;$%
and $V_{h_{\pm }}\left( z\right) $ in (81-83) are functions of $z^{p}.\;$As
a consequence, the wave functions (and the Green functions) become
multivalued; however, this transformation allows a reinterpretation of the
results in terms of twisted conformal field theory$^{\left[ 23\right] }$. In
this method the fields $\phi _{s}\left( z\right) $ are introduced by hands
using the properties of conformal field theory; in our method the fields
come naturally from Calogero-Sutherland and Ruijsenaars-Schneider theory via
the specific limit $q\rightarrow \exp (2i\pi /p)$ which fits perfectly well
the idea of ''composite particles''. In Ref$\left[ 22\right] $, the authors
conclude that ''any non-null wave function can be written as cluster of $p$
one-electrons fields''; clearly, our approach runs the other way round, from
''composite particles'' to conformal field theory.

\bigskip

\bigskip

\bigskip

\bigskip {\Large Acknowledgments}

\bigskip I wish to thank M.\ Bauer for his useful comments and for a careful
reading of the manuscript. I also thank the referee who suggested to
introduce the quasihole operator.

\bigskip

\bigskip

$\left[ 1\right] \;\ \ $M.C. Berg\`{e}re, Fractional statistic,
cond-mat/9904227.

$\left[ 2\right] \;$\ \ F.D.M. Haldane, Phys.\ Rev.\ Lett. {\bf 67}, 937
(1991).

$\left[ 3\right] $ \ \ M.C.\ Berg\`{e}re, The partition function for
''composite particles'',

\ \ \ \ \ \ \ cond-mat/9909179.

$\left[ 4\right] $\ \ \ C.N. Yang and C.P.\ Yang, J.\ Math.\ Phys. {\bf 10},
1115 (1969).

$\left[ 5\right] \;$ \ B.\ Sutherland, J.\ Math.\ Phys. {\bf 12}, 251
(1971); Phys. Rev. {\bf A4}, 2019

\ \ \ \ \ \ \ (1971); {\bf A5}, 1372 (1972).

$\left[ 6\right] $ \ \ Y.S.\ Wu, Phys.\ Rev.\ Lett. {\bf 73}, 922 (1994).

$\left[ 7\right] $ \ \ D.\ Bernard, Les Houches Session LXII (1994).

$\left[ 8\right] $ \ \ A.\ Dasni\`{e}re de Veigy and S.\ Ouvry, Phys.\ Rev.\
Lett. {\bf 72}, 600 (1994).

$\left[ 9\right] $ \ \ K.\ Hikami, Phys.\ Lett.\ {\bf A205}, 364 (1995).

$\left[ 10\right] \ $R. van Elburg and K.\ Schoutens, Phys.\ Rev. {\bf B58},
15704 (1998).

$\left[ 11\right] $ R.\ Kedem, T.R.\ Klassen, B.M.\ McCoy and E.\ Melzer,
Phys.\ Lett. {\bf B304},

\ \ \ \ \ \ 263\ (1993); {\bf B307}, 68 (1993).

$\left[ 12\right] $ A.\ Berkovitch and B.M.\ McCoy, The universal chiral
partition function

\ \ \ \ \ \ for\ exclusion statistics, hep-th/9808013.

$\left[ 13\right] $ F\ Calogero, J.\ Math.\ Phys. {\bf 10}, 2191, 2197
(1969).

$\left[ 14\right] $ S.N.M. Ruijsenaars and H.\ Schneider, Ann. Phys. {\bf 170%
}, 370 (1986);

\ \ \ \ \ \ S.N.M\ Ruijsenaars, Comm. Math.\ Phys. {\bf 110}, 191 (1987).

$\left[ 15\right] $ R.B.\ Laughlin, Phys.\ Rev.\ Lett. {\bf 50}, 1395 (1983).

$\left[ 16\right] $ J.K.\ Jain, Phys. Rev.\ Lett. {\bf 63}, 199 (1989);
Phys.\ Rev. {\bf B41}, 7653

\ \ \ \ \ \ (1990); Adv. in Phys. {\bf 41}, 105 (1992).

$\left[ 17\right] \;$H.A.\ Bethe, Zeitschrift f\"{u}r Physik, {\bf 71}, 205
(1931).

\ \ \ \ \ \ M.\ Gaudin, La fonction d'onde de Bethe, Masson (1983).

$\left[ 18\right] $ Y.\ Asai, M.\ Jimbo, T.\ Miwa and Y. Pugai, J.\ Phys. A:
Math.\ Gen. {\bf 29},

\ \ \ \ \ \ 6595, (1996); S.L. Lukyanov and Y.\ Pugai, Nucl.\ Phys. {\bf B473%
}, 631

\ \ \ \ \ \ (1996); JETP, {\bf 82}, 1021 (1996); H.\ Awata, H.\ Kubo, S.\
Odake and

\ \ \ \ \ \ J. Shiraishi, Com. Math.\ Phys, {\bf 179}, 401 (1996).

$\left[ 19\right] $ I.G. Macdonald, Symmetric functions and Hall
polynomials, 2nd ed.

\ \ \ \ \ \ Clarendon Press (1995).

$\left[ 20\right] \;$D. Uglov, Com.\ Math.\ Phys. {\bf 191}, 663 (1998).\ \ 

$\left[ 21\right] $\ V.\ Pasquier and D.\ Serban, CFT and edge excitations
for the principal

\ \ \ \ \ \ \ series of quantum Hall fluids, cond-mat/9912218.

$\left[ 22\right] $ G.\ Cristofano, G.\ Maiella and V.\ Marotta, A twisted
conformal field

\ \ \ \ \ \ \ theory\ description of the quantum Hall effect,
cond-mat/9912287.

$\left[ 23\right] \;$M.\ Bershadsky and A.\ Radul, Int.\ J.\ Mod.\ Phys. A%
{\bf 2}, 165 (1987).

\ \ \ \ \ \ \ L.\ Dixon, D. Friedan, E. Martinec and S.\ Shenker, Nucl.\
Phys. B{\bf 282},

\ \ \ \ \ \ \ 13 (1987).

$\left[ 24\right] $ \ X.\ G.\ Wen, Int.\ J. of Mod. Phys. {\bf B6}, 1711
(1992).\ \ \ \ \ \ \ \ \qquad\ \ \ \ \ \ \ \ 

\newpage
\begin{figure*}
\centerline{\psfig{figure=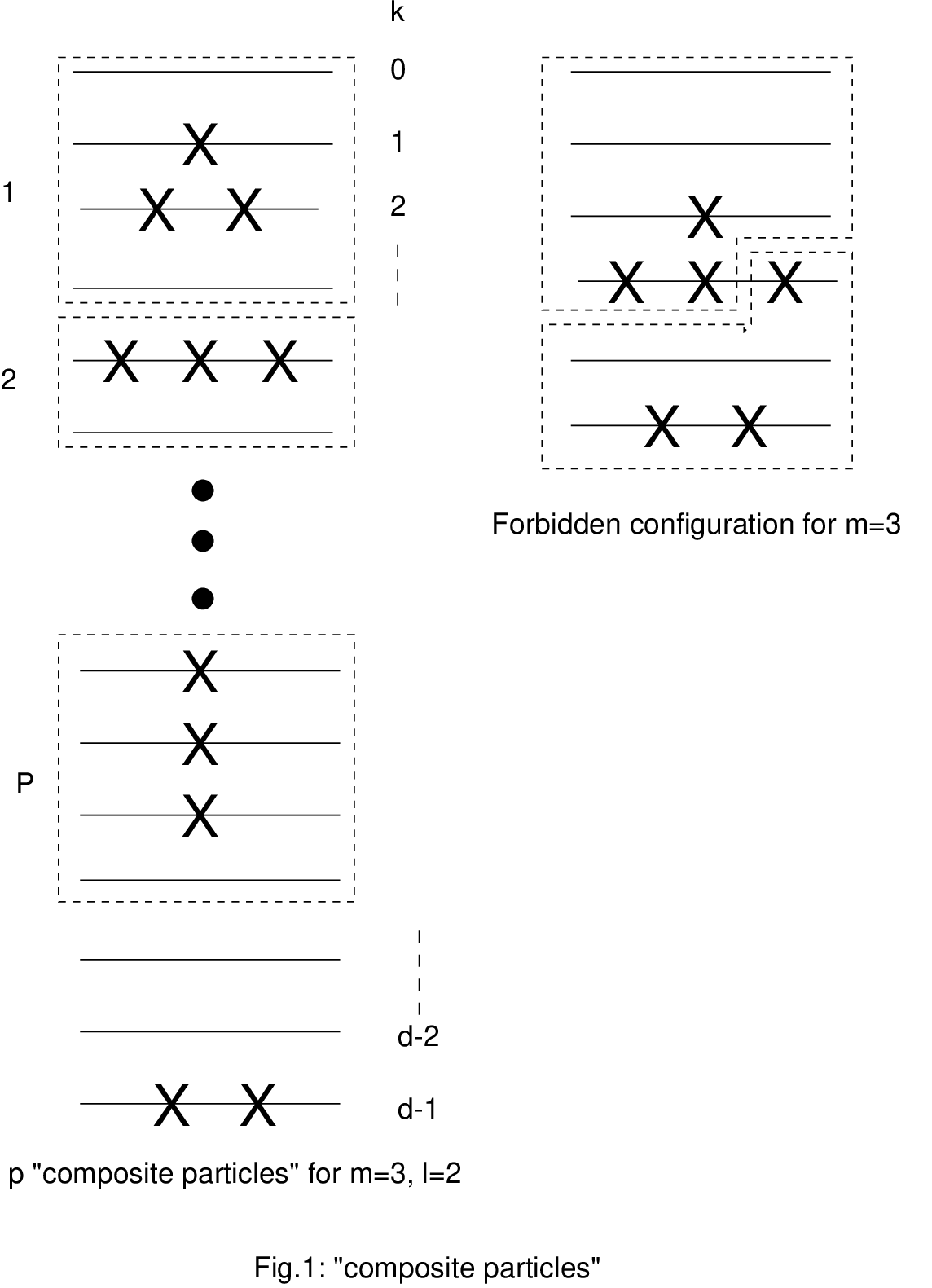,width=15cm,height=15cm}}
\end{figure*}
\newpage
\begin{figure*}
\centerline{\psfig{figure=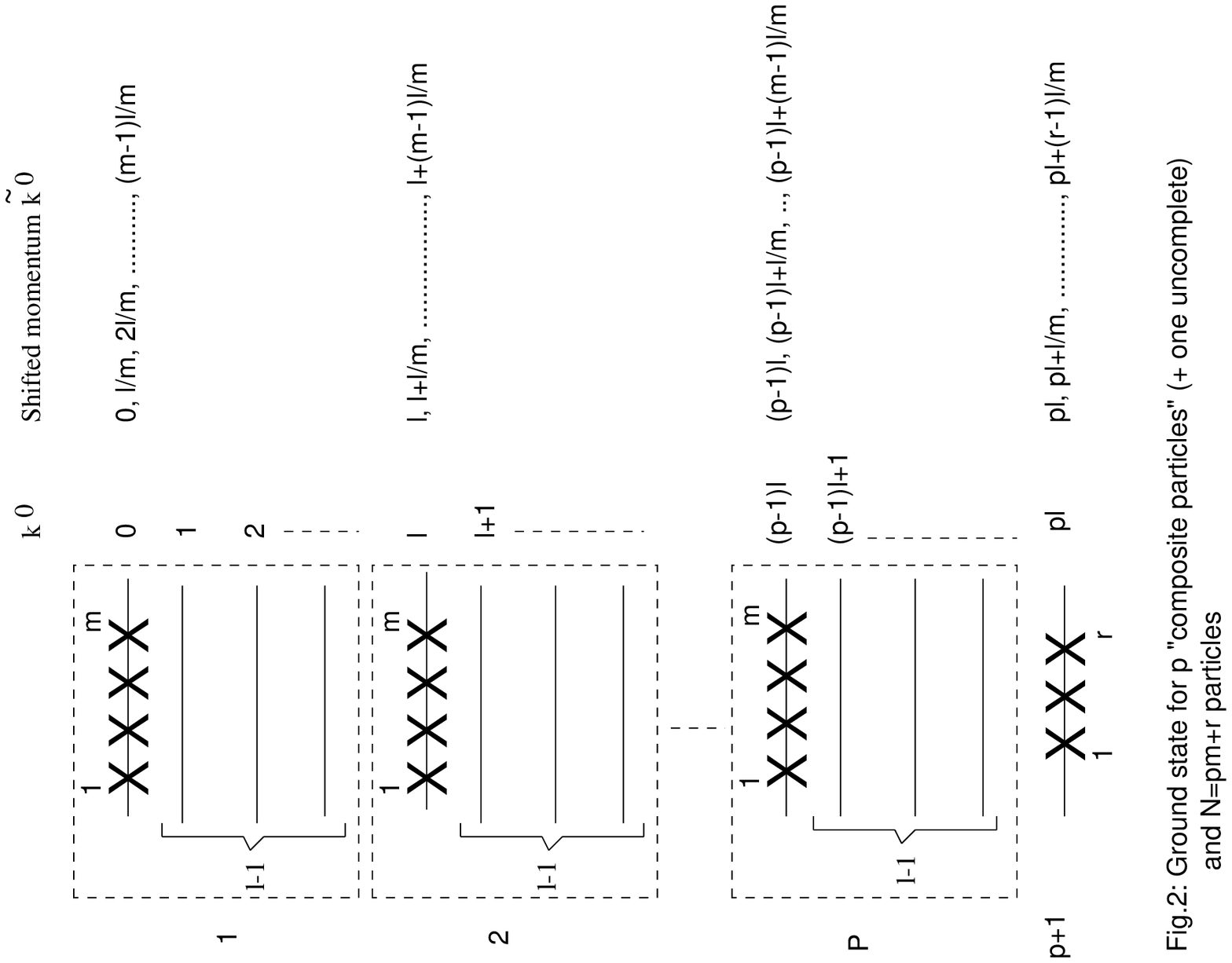,width=15cm,height=15cm}}
\end{figure*}
\begin{figure*}
\centerline{\psfig{figure=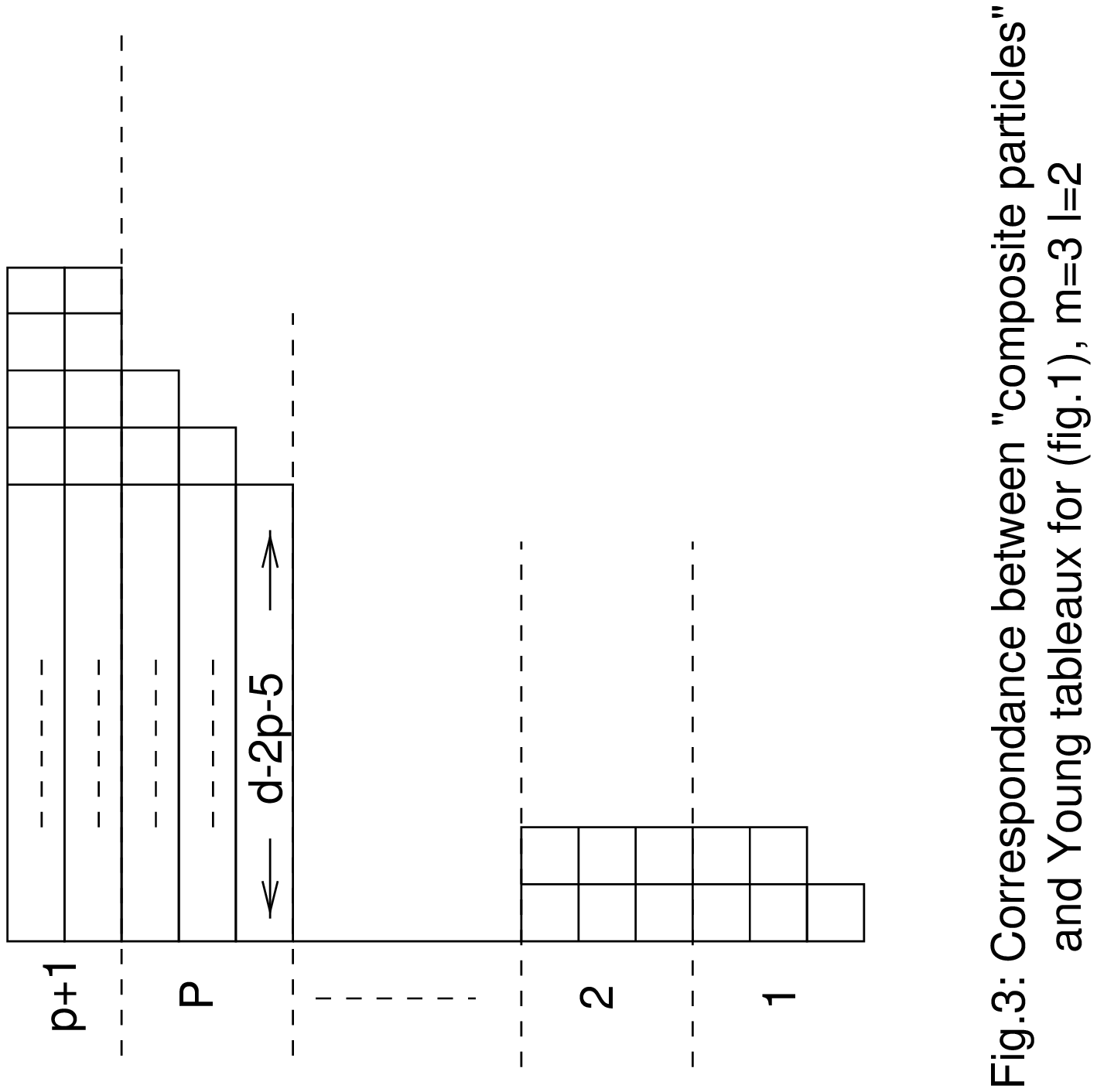,width=15cm,height=15cm}}
\end{figure*}
\begin{figure*}
\centerline{\psfig{figure=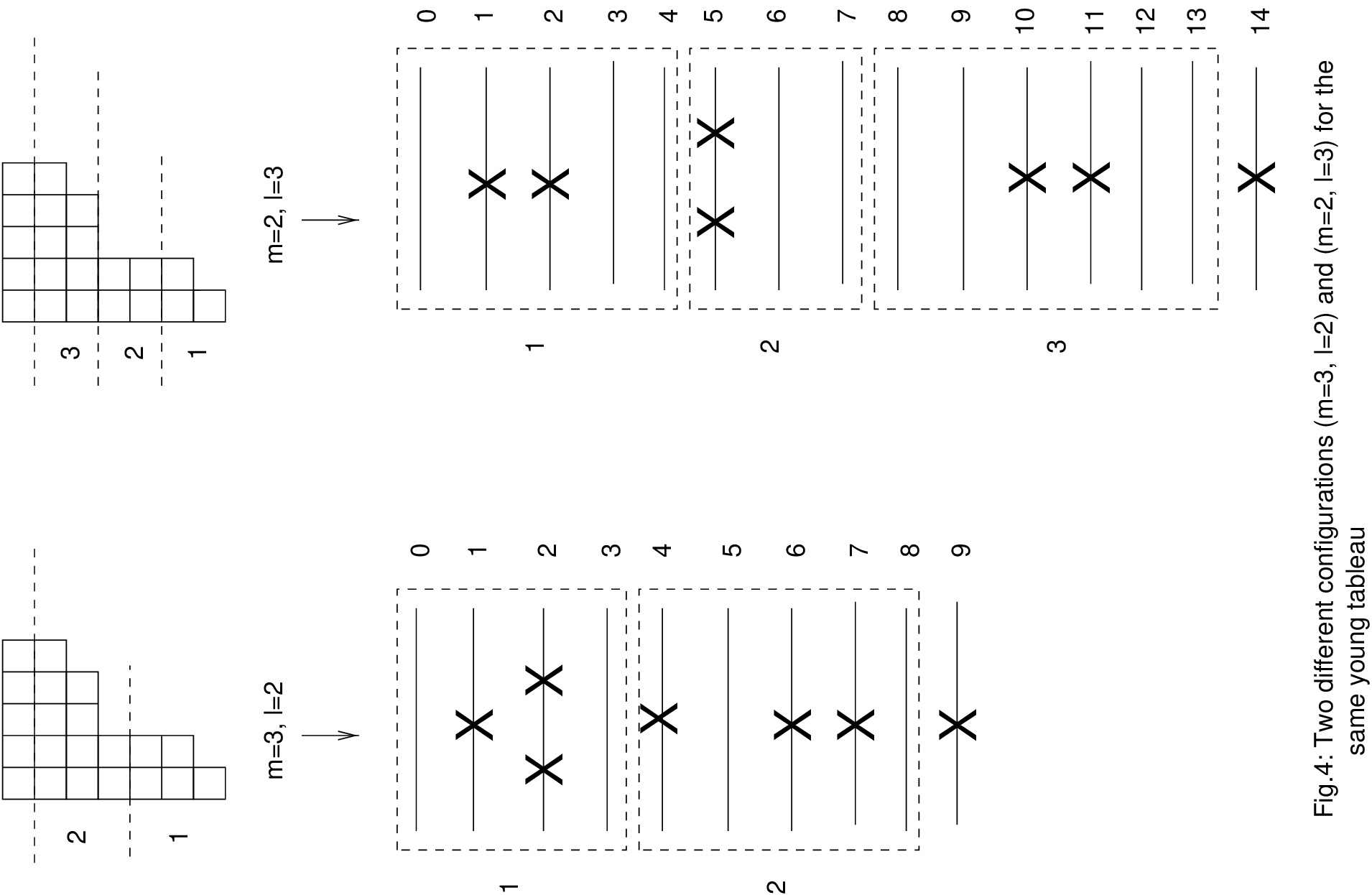,height=15cm,width=15cm}}
\end{figure*}
\end{document}